# Unsupervised Machine Learning Discovery of Chemical and Physical Transformation Pathways from Imaging Data

Sergei V. Kalinin,[1,a] Ondrej Dyck,[1] Ayana Ghosh,[1,2] Yongtao Liu,[1] Roger Proksch,[3] Bobby G. Sumpter,[1] and Maxim Ziatdinov[1,2,b]

[1] Center for Nanophase Materials Sciences and [2] Computational Sciences and Engineering Division, Oak Ridge National Laboratory, Oak Ridge, TN 3783, United States

[3] Asylum Research, Oxford Instruments, 6310 Hollister Ave, Goleta, CA 93117, United States

We show that unsupervised machine learning can be used to learn physical and chemical transformation pathways from the observational microscopic data, as demonstrated for atomically resolved images in Scanning Transmission Electron Microscopy (STEM) and ferroelectric domain structures in Piezoresponse Force Microscopy (PFM). To enable this analysis in STEM, we assumed the existence of atoms, a discreteness of atomic classes, and the presence of an explicit relationship between the observed STEM contrast and the presence of atomic units. In PFM, we assumed the uniquely-defined relationship between the measured signal and polarization distribution. With only these postulates, we developed a machine learning method leveraging a rotationally-invariant variational autoencoder (rVAE) that can identify the existing structural units observed within a material. The approach encodes the information contained in image sequences using a small number of latent variables, allowing the exploration of chemical and physical transformation pathways via the latent space of the system. The results suggest that the high-veracity imaging data can be used to derive fundamental physical and chemical mechanisms involved, by providing encodings of the observed structures that act as bottom-up equivalents of structural order parameters. The approach also demonstrates the potential of variational (i.e., Bayesian) methods for physical sciences and will stimulate the development of new ways to encode physical constraints in the encoder-decoder architectures, and generative physical laws, topological invariances, and causal relationships in the latent space of VAEs.

[a] sergei2@ornl.gov
[b] ziatdinovma@ornl.gov

Over the last decade, deep machine learning has become a key enabling technology in multiple areas of computer science, imaging, and robotics.[1-5] Following the successful demonstration of deep convolutional networks for image recognition tasks,[6] deep learning architectures have helped revolutionize many other areas, ranging from natural language processing, reinforcement learning for control,[7-9] and have begun to rapidly propagate in areas such as causal and physical discovery,[10-12] automated experiments in chemistry, materials science, and biology,[13-15] and direct atomic manipulation and assembly.[16, 17]

This rapid growth in machine learning (ML) applications has naturally attracted the attention of domain science communities for the potential application of deep learning to scientific discovery.[18-20] While immediately attractive, the examination of this concept suggests that this is far from trivial. Indeed, it is by now well understood that classical ML methods are correlative in nature and serve as powerful interpolators operating within distributions spanned by training data.[21] Through the choice of the training data or network architecture, the equivariance or symmetries can be introduced, imposing physical constraints on the derived outputs. However, the capability of deep neural networks (DNNs) to extrapolate outside of the training domain or work with out-of-distribution data represents a major limitation.[22, 23] This is in comparison to the classical scientific research paradigm, often relying on past knowledge, deductive logic based on known physical laws, causal physical mechanisms, and hypothesis driven paradigms. As such, scientific discovery in the physical sciences typically relies on relatively small amounts of data and allows extrapolation well outside the specific domains, which are the "hard" tasks for traditional ML. Thus, the merger of ML and classical scientific methods may provide a new forefront for research and development.

One such approach is based on the introduction of known physical constraints in the form of symmetries, conservation laws, etc., to limit ML prediction to the physically possible. Simple examples of this approach are use of sparsity, non-negativity, or sum to one constraint in linear and nonlinear unmixing,[24] resembling the classical Lagrange multiplier approach. Several groups have introduced the physical constraints of Hamiltonian mechanics in the deep neural networks, significantly improving predictions of long-term dynamics in chaotic systems.[25] Similarly, the network architecture can be made to encode known symmetry constraints, enabling the discovery of order parameters.[26] Alternatively, constraints can be introduced on the form and complexity of structural and generative behaviors, as exemplified by the combination of symbolic regressions and genetic algorithms.[27, 28] However, until now most of these developments have been limited to areas such as mechanics and astronomy, where the generative laws are relatively simple and well-defined and hence ML predictions can be readily compared to physical models.

At the same time, discovery of more complex laws and reduced rules that define fields of chemistry, molecular biology, etc., have not yet been demonstrated. In these fields, the elementary descriptors of the systems are generally not well defined, and the constraints imposed on the functional relationships between the descriptors are generally "soft". Combined with the preponderance of large data sets, this led to broad development of the correlative ML models. Interestingly, even causal machine learning methods first emerged in the context of economics or biology, rather than physical sciences. Correspondingly, the open question remains to what extent can we use machine learning for discovery in specific domain areas, given the experimental observations and minimal set of prior knowledge in the form of postulated descriptors or generative mechanisms.

Here, we demonstrate how unsupervised machine learning can be used to discover elementary building blocks forming the structure of solids, illustrated by the structural units and chemical transformation mechanisms in defected graphene from atomically resolved scanning transmission electron microscopy (STEM) data and ferroelectric domain wall dynamics from the Pizoresponse Force Microscopy (PFM) data. The postulates that we repeatedly use during the workflow is that of the existence and observability of atoms in STEM and relationship between measured contrast and ferroelectric walls in PFM. We show this is sufficient to discover molecular structures and map chemical transformation mechanisms in graphene and domain wall dynamics in PFM. While for STEM these results are directly comparable to the known tropes of organic chemistry and physics, a similar approach can be extended to the areas where the domain knowledge is less developed or non-existent as illustrated by PFM example.

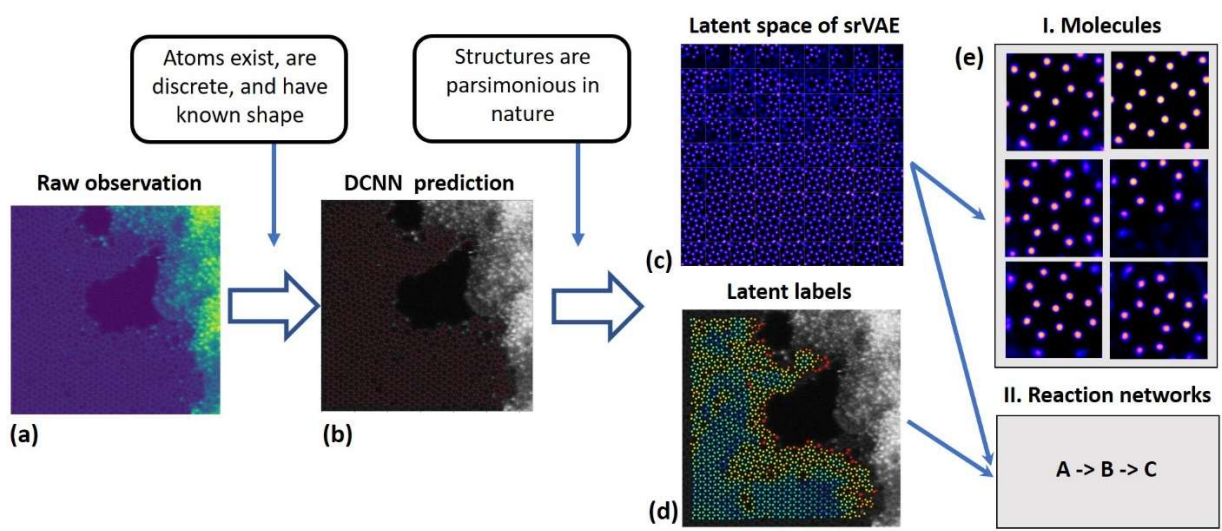

**Figure 1.** Discovery of molecular structural fragments and chemical transformation mechanisms via unsupervised machine learning. Here, a deep convolutional neural network (DCNN) is used to perform the semantic segmentation of (a) the experimental data set to yield (b) the coordinates and identities of atomic species. Note that for the "ideal" experimental data, similar results can be achieved using the simple maxima analysis/blob finding, and a DCNN is used to visualize atoms only. Further shown are the (c) latent space of the rotationally-invariant variational autoencoder with skip connectivity between the latent space and decoder (srVAE) and (d) image encoded with the one of the srVAE latent variables. (e) The analysis of the latent space of the srVAE illustrates the regions containing the easily identifiable molecular fragments, e.g., 6- and 5- member cycles, 5-7 defects, and edge configurations. Note that these are discovered in a fully unsupervised manner. Comparing the evolution of latent variables corresponding to a single atom further allows discovery of chemical transformation mechanisms.

As one model system, we have explored the rich set of electron beam (e-beam) induced transformations in Si-containing graphene. This system has been shown to possess a broad range of e-beam induced transformations, including formation of point and extended defects, migration of Si atoms, emergence of small-angle boundaries and fragmentation of host graphene lattice, and eventually formation of defect clusters and degradation.[29-43] Sample preparation details are provided in the methods section. Figure 1 (a) shows a region of the specimen where suspended clean graphene was observed next to an amorphous region containing Si and Cr atoms. The electron beam was scanned over the field of view recording a medium angle annular dark field (MAADF) image sequence during irradiation and inducing slow chemical changes in the suspended graphene. Here, we aim to explore whether the elementary bonding patterns (chemical fragments) in this system and their changes with time (chemical transformations) can be derived from the STEM observations in an unsupervised manner with a minimum number of an *a priori* assumptions and postulates. In other words, can we create a computational architecture that can discover chemistry and chemical transformation pathways from observations in an unsupervised manner? While in this case the results can be compared with prior chemical knowledge in a straightforward fashion, similar approaches applied to unknown systems will offer a pathway to scientific discovery.

Our approach is based on the concept of a variational autoencoder (VAE)[44] which finds the latent representation of the complex system. Generally, a VAE is a directed latent-variable probabilistic graphical model that learns a stochastic mapping between observations $x$ with a complicated empirical distribution and latent variables $z$ whose distribution can be relatively simple.[45] A VAE consists of a generative model ("decoder") that reconstructs $x_i$ from a latent "code" $z_i$ and an inference model ("encoder") whose role is to approximate a posterior of the generative model via amortized variational inference.[46] Implementation-wise, both encoder and decoder models are approximated by deep neural networks whose parameters are learned jointly by maximizing the evidence lower boundary via a stochastic gradient descent with randomly drawn mini-batches of data. In this manner, VAEs allow one to build relationships between high-dimensional data sets and a small number of latent variables, somewhat reminiscent of manifold learning and self-organized feature maps.

The important aspect of the VAEs (similar to many manifold learning methods) is that the variability of the behaviors in the latent space allows one to reveal relevant features of the system behavior, i.e., primary non-linear degrees of freedom. One example of this is the hand gesture analysis. Another is writing styles or emotional expressions. This capability of VAEs to disentangle data representations has attracted broad attention from the computer vision community and is being explored here. Second an important aspect of VAEs employed here is the principle of parsimony – i.e., the training process generates the best short descriptors representing the data, somewhat reminiscent of Occam's principle.

However, direct application of the VAE to experimental data requires two special aspects. One is a suitable choice of the raw descriptors, namely suitably chosen sub-sets of the images (as compared to classical applications such as analysis of MNIST or CIFAR data sets, where images are used in full). Establishing these descriptors requires prior knowledge of the system, since randomly picking sub-images (zero prior knowledge) tends to lead to suboptimal results. Here as the prior knowledge, we use the most basic assumptions, i.e., the existence of atoms, discreteness of atom classes, and discoverability of atoms from STEM data. For the latter, we implicitly rely on the fact that the maximum of the STEM contrast corresponds to the location of an atomic nucleus and the intensity is proportional to the atomic number, i.e., it enables identification of the atoms and their types. For PFM, we assume the high veracity, that is, free of observational biases and confounding factors, relationship between the signal and the presence of the ferroelectric domain walls.

To implement this for STEM, we used deep convolutional neural networks (DCNN) trained first on simulated data and then iteratively retrained to adapt to experimental images. The details about training DCNNs on simulated/synthetic atom-resolved data were reported earlier by multiple groups, including the authors of this study.[47, 48] Here, we found that when a DCNN trained on simulated data is applied to the experimental data characterized by rapidly growing amorphous regions and holes in the lattice (up to ~60 % (combined) of the entire image towards the end of the movie) works well only at the beginning of the movie, and its predictions rapidly deteriorate as the system disorder increases. For example, it produces a large number of false positives inside the graphene holes. This behavior is not surprising and is in fact expected when applying a deep learning model to out-of-distribution data.[23] To address this issue, we used the DCNN predictions on the first ~5 experimental frames where detected atomic positions can be refined via standard Gaussian fitting to create a new training set, then retrained a model and applied it to the entire movie. The re-trained model demonstrated a significant improvement in the detection rate of atomic species and, most importantly, was not prone to artifacts that plagued the performance of the initial model toward the end of the movie.

The application of the DCNN to the raw experimental data allows transformation of the STEM image stack into the semantically segmented data set giving the probability that a specific pixel belongs to an atom. Combined with the simple thresholding and blob finding, this allows straightforward and highly robust translation of raw STEM data into the atomic coordinates and identities. Note that while DCNNs are a supervised learning technique, they are trained on the labeled data sets that postulate only the existence of atomic species and define how they manifest in the STEM data. However, there are no assumptions on how atomic units are connected and how these connectivities evolve with time (i.e., chemical reactivity).

The DCNN analysis of the STEM images allows decoding of the atomic coordinates and their evolution with time. Note that similar information can be derived from e.g., theoretical modeling or "ideal" imaging with very high signal/noise ratio, bypassing the DCNN step. Furthermore, it is important to note that experimentally the carbon atoms are difficult to distinguish, i.e., trajectories cannot be easily reconstructed. However, the well-separated Si atoms allow convenient markers without loss of generality of the approach. With the atomic positions (for C and Si) identified, we form the experimental descriptors as the sub-images of a given size, *N*, centered on atomic units. These descriptors combine the knowledge of atom existence (coordinates) and experimental data (raw or semantically segmented contrast). However, they do not contain any prior information on chemical bonding or larger level structural blocks.

To discover salient features of system behavior from the bottom up, we need the latent representation of the data that can be obtained using a VAE. We note that the possible chemical building blocks can have different orientations within the image, necessitating development of the VAE architecture invariant with respect to rotations. Here, we adapted the rotationally invariant VAE (rVAE) originally proposed by Bepler et al.[49] and adapted for analysis of dynamic scanning probe and transmission electron microscopy data by Ziatdinov et al.[50, 51] We note that the shortcoming of the traditional VAE's encoder-decoder architectures is the so-called posterior collapse,[52] *i.e.,* when the posterior estimates of a latent variable $z_i$ do not provide a good representation of the data. To alleviate this problem and emphasize the reconstruction rather than encoding of the data, here we connected the latent space with each layer of the rVAE's decoder's neural network via the skip connections,[53, 54] thereby enforcing a dependence between the observations and their latent variables. We note that this approach is different from a well-known method of adding residual ("skip") connections between different layers of a neural network[55] since in our case the connection paths originate in the latent space.

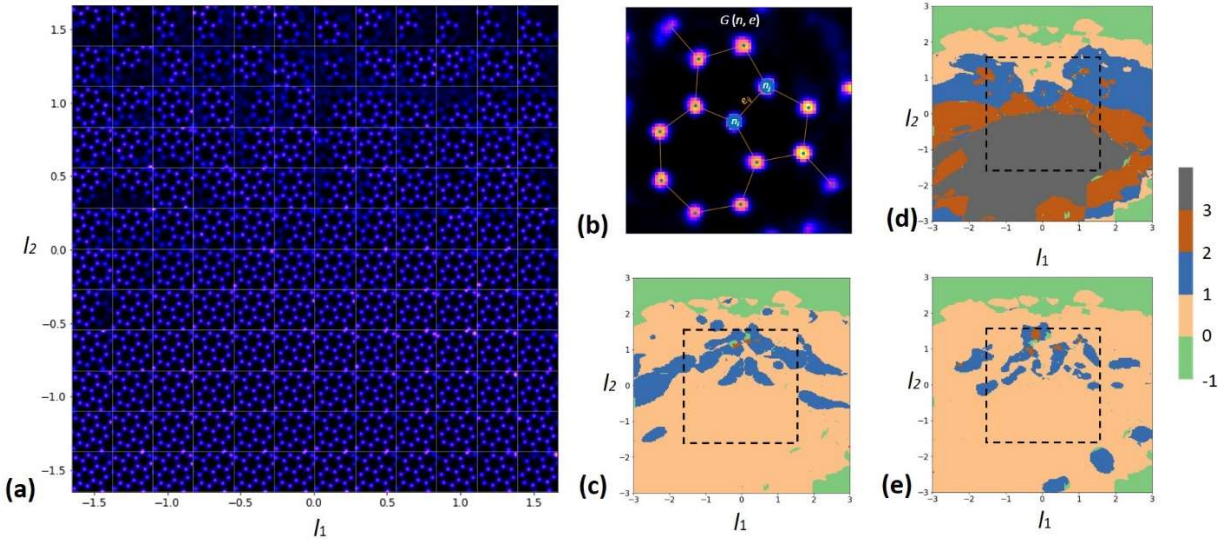

**Figure 2.** Latent space analysis for atom-centered descriptors. (a) Latent space of the skip-rVAE for low sampling (12x12). (b) The graph analysis of the sub-image corresponding to a selected point in the latent space. Distributions of the (c) pentagonal, (d) hexagonal, and (e) heptagonal rings adjacent to the central atom in the latent space for high sampling (200x200). Here, -1 corresponds to unphysical configurations and 1, 2, and 3 are for number of rings of particular type surrounding central atom. The dotted boxes in (c-e) correspond to latent space area in (a).

The latent space of the skip-rVAE trained on the experimental data is shown in Figure 2 (a). Here, the rectangular grid of points $l_1 \in [l_{1min}, l_{1max}]$, $l_2 \in [l_{2min}, l_{2max}]$ is formed and the images formed by decoding the chosen ($l_1$, $l_2$) pair are plotted on a rectangular grid. This depiction allows observation of the evolution of sub-images in the latent space and establishes the relationship between the STEM image (i.e., local atomic structure) and the latent variables. Note that not all combinations of latent variables correspond to physically possible STEM images, and hence distributions of the experimental data in the latent space can have a complicated structure. Similarly, decoded images need to be ascertained for physicality.

Here, we explore whether latent space reconstructions contain information on the molecular building blocks in the graphene lattice. By construction of descriptors, the center of a reconstructed sub-image will contain a single well-defined atom (Si or C). Hence, the remainder of the image contrast can be directly interpreted in terms of whether atomic structures (as opposed to some abstract representations) are observed, and what these structures are. Surprisingly, casual examination of Fig. 2 (a) illustrates that even at a low sampling density, the latent space representations generally correspond to well-defined molecular graphene fragments, comporting to the chemical intuition of an organic chemist. The lower half of the depicted latent space is comprised of the structures formed by the three 6-membered rings, an elementary building block of graphene for this window size. Structures in the upper part of the diagram illustrate the presence of edges, or rings with different numbers of members. A part of the latent space contains well-known 5-7 defects (magnified in Fig. 2 (b)). The "unphysical" images with smeared or weak contrast manifest in only a few locations and are an unalienable feature of the projection of discrete system on low dimensional manifold.

To gain further insight into the system behavior in the latent space and establish relationships between latent variables and classical organic chemistry descriptors, we classify the observed structures based on connectivity of the carbon network. Here, we developed a simple approach where all units above an intensity threshold $t = 0.5$ in the images projected from [$l_1$, $l_2$] coordinates of the latent space, are considered to be physical. We then identified a graph structure $G(n, E)$ that corresponds to these atomic units, where $n$ is number of nodes (units) in the image and $E$ corresponds to edges connecting different nodes. Only nodes separated by $0.5 l_{C-C} < d < 1.2 l_{C-C}$ are defined as connected. The lower bound is set such that it can potentially account for out-of-plane distortions which results in the apparent shrinkage of bonds as seen from the 2D projection in STEM. We note that such an analysis works only for 2D systems. Finally, a depth-first search method is used to traverse a graph structure and identify a number of different $n$-membered rings adjacent to the central atom. Note that this approach can be extended further to explore additional details of the atomic structure beyond adjacent rings (e.g., broken bonds, dangling atoms, etc.), but this analysis will require larger window sizes and will come at the cost of simplicity.

The distribution of the number of 5-, 6- and 7- member rings in the latent space calculated with a high sampling is shown in Figures 2 (c-e). The top side of the images (positive $l_2$) generally contains a small number of rings and corresponds to edges or isolated atoms. The different values of the latent variables encode structural deformations within graphene. The bulk of $l_2 < 0$ region corresponds to a normal graphene structure. Finally, the region for $-1 < l_1 < 1$ and $0 < l_2 < 1.5$ contains islands with varying number of the 5- and 7-member rings. Interestingly, overlap of these islands corresponds to the 5-7 defects.

Explicit examination of the images in the different parts of the latent space as well as the morphologies of the domains in Fig. 2 (c-e) suggests that the skip-rVAE has discovered the latent representations of the observed STEM contrast in terms of continuous latent variables, and these representations separate the possible structures of chemical bonding networks. The well-defined chemical structures occupy specific regions of the latent space with relative areas proportional to the fraction of these structures in the initial data set. At higher sampling densities, these regions are often separated by the thin lines of "unphysical" domains corresponding to the transitions between physically realizable configurations via "impossible" configurations such as smeared atoms, etc. Finally, within each uniform region the structure variability along the lattice space represents the physical distortions or structure outside the primary rings.

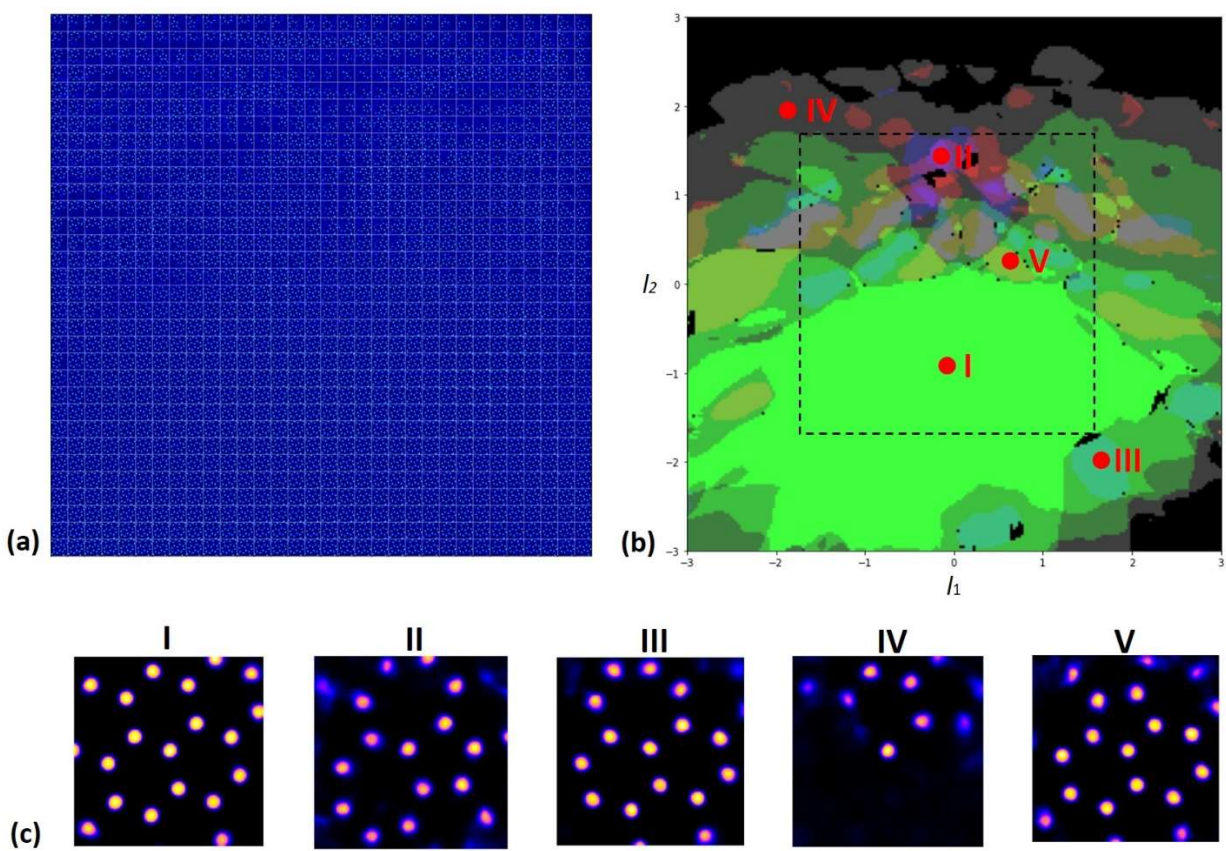

**Figure 3.** Chemical space analysis for atom-centered descriptors. (a) Latent space of skip-rVAE at high sampling. (b) Chemical space map and (c) several examples of the observed structures, including (I) prototype graphene, (II) 5-7-7 defect, (III) 7- member ring, (IV) fragment of bearded edge, and (V) 5-member ring. The dotted box in (b) corresponds to the area shown in (a).

These behaviors are further illustrated in Figure 3. Figure 3 (a) shows the latent space of the system at high sampling. While the details of individual images cannot be discerned, the overall smooth evolution of decoded patterns is clearly visible, as are large scale variations in encoded behaviors. Figure 3 (b) represents the overlay of Figures 2 (c-e), visualizing the domains with dissimilar chemical structures. Finally, several structures corresponding to selected regions of the latent space are shown in Figure 3 (c), including (I) prototype graphene, (II) 5-7-7 defect, (III) 7-member rings, (IV) edge region, and (V) 5-member ring. We note that the complexity of chemical space increases with the size of sub-image descriptors since, e.g., a 5-member ring can be adjusted to 6-, 7- and 8- membered-rings in different realizations of topological defects.

To simplify the representation of chemical space we also created an alternative set of descriptors for skip-rVAE training where we used sub-images centered on hollow sites instead of atoms. This is achieved by applying the graph analysis described above to the DCNN output from the entire dataset and computing the center of the mass for each identified cycle (with the maximum cycle length set to 8). By design, this description is suited for analyzing the material microstructure on the level of single rings. It is again important to mention that the information available to the network is the positions of the centers and patches of the image centered at these positions; rather than atomic coordinates or any of the higher-level descriptors (bonds, orientations etc.). The skip-rVAE results for the new set of descriptors are shown in Figure 4. Here, both the latent manifold (Figure 4a) and the chemical space map (Figure 4b) have a much simpler structure exhibiting well-defined regions associated with 5, 6, 7- and 8-member rings (Figure 4c).

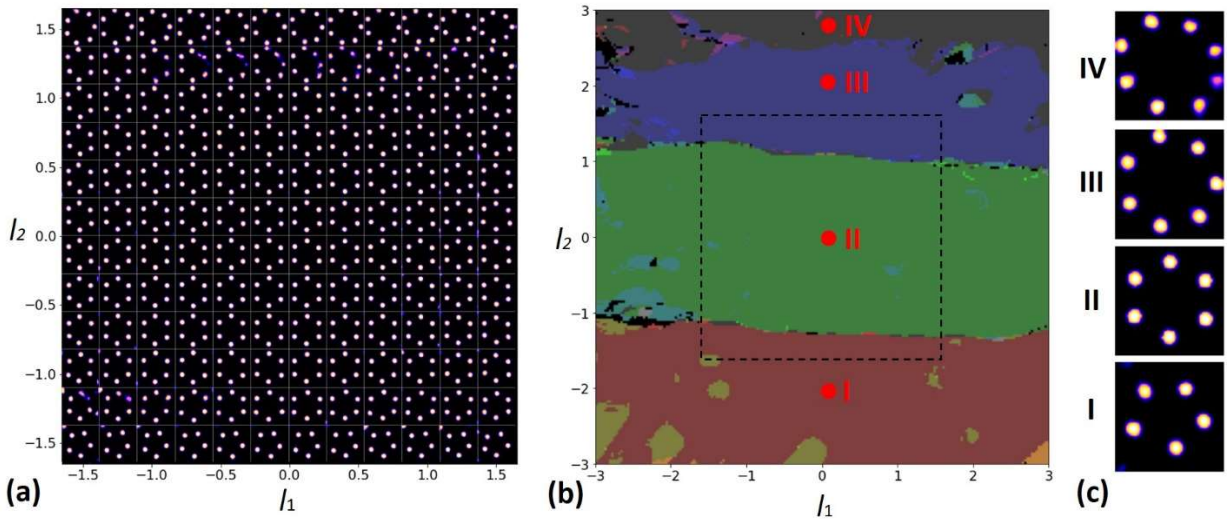

**Figure 4.** Chemical space analysis for hollow-site-centered descriptors. (a) Latent space of skip-rVAE for 12x12 sampling. (b) Chemical space map and (c) examples of the ring structures from each of the four well-defined regions on the map in (b). The dotted box in (b) corresponds to the area shown in (a).

This analysis naturally leads to a question as to whether a different dimensionality of the latent space can be chosen. From a general perspective, finite discrete space (i.e., all arrangements of carbon atoms containing no more than a certain number) can be projected even onto the 1D continuous distribution. Hence, for the discrete data (possible structural fragments) the encoding can be performed by 1, 2, or higher dimensional continuous variables. However, for 1D the encoding will require a much larger number of significant digits, whereas 3- and higher dimensional latent spaces do not allow for straightforward visualization similar to Fig. 3 (a). Hence, we chose 2D latent spaces for convenience, and note that a similar approach is used in many other branches of ML, e.g., in analysis of self-organized feature maps (SOFMs) or graph embeddings.

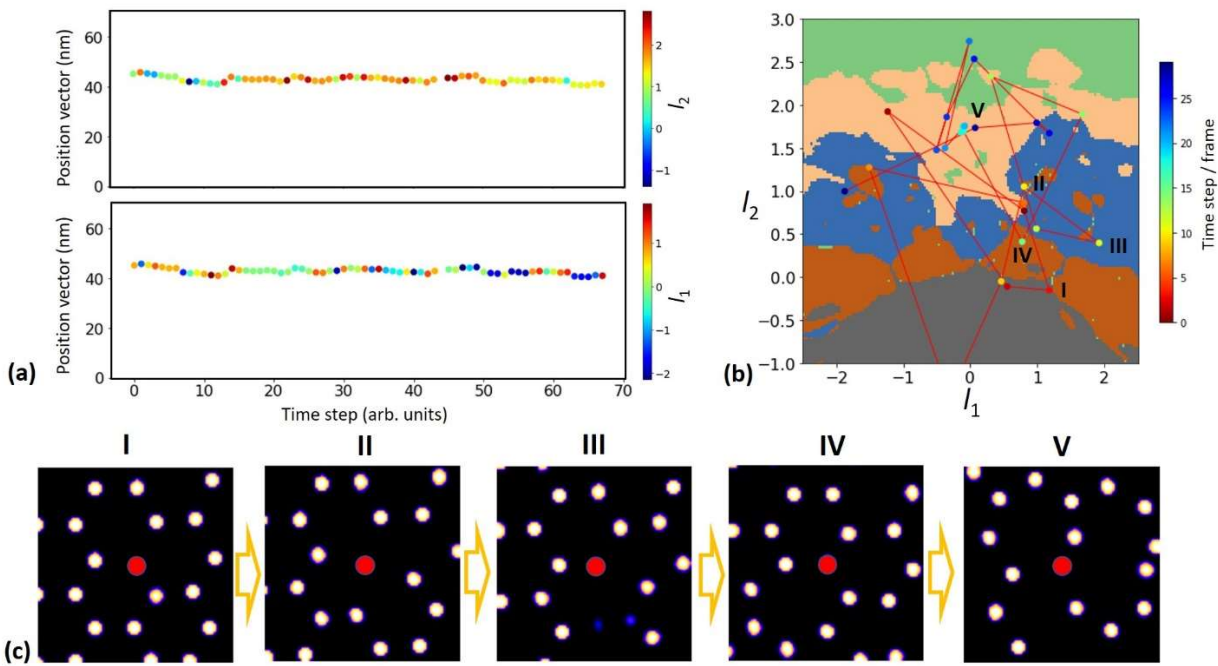

**Figure 5.** Chemical evolution of a selected Si atom during e-beam irradiation. (a) Evolution of the latent variables as a function of time. The position vector is calculated as $\sqrt{x^2 + y^2}$ from the image origin. (b) dynamics on the latent plane represented as a map of hexagonal rings (see Figure 3d) for the first 30 frames (the scatter plot for all the time steps is available from the accompanying notebook). (c) Chemical neighborhoods during the beam-induced transformations including transition from (I) 3-fold coordinated to (II) 4- fold coordinated Si, to (III) transitional structure to (IV) 3-fold coordinated graphene with adjacent 7-ring to (V) more complex patterns involving formations of quasi-linear chains. Note that central atom is always Si. The depicted structures correspond to frames 3, 10, 11, 15 and 28 in (a).

Finally, this approach allows exploring the chemical dynamics, i.e., the transformations of the chemical bonding network during the e-beam irradiated process. Generally, this necessitates tracing of the atoms between the frames and reconstruction of the trajectories. However, in this case significant changes in the carbon network between the frames makes it difficult, especially for atoms where the bonding changes. However, Si atoms that are present in the graphene lattice offer readily identifiable markers that can be often traced between frames. An example of such an analysis is shown in Figure 5. Figure 5 (a) depicts the evolution of latent variables for a selected Si atom through the ~70 frames. Initially, the $l_2$ values are close to 0, corresponding to the Si on a three-coordinated lattice site, i.e., a substitutional defect. In the initial stage of the process, the atom moves within the "ideal" graphene regions, with the changes in latent variables representing the changes in strain state and distant neighborhood. The 3-fold Si can transform into 4-fold Si (I-II in Figure 5c), clearly visible in the latent space of the system (Figure 5b). Subsequently, more complex coordinations can emerge, as visualized in states (III-V) in Figure 5 (c).

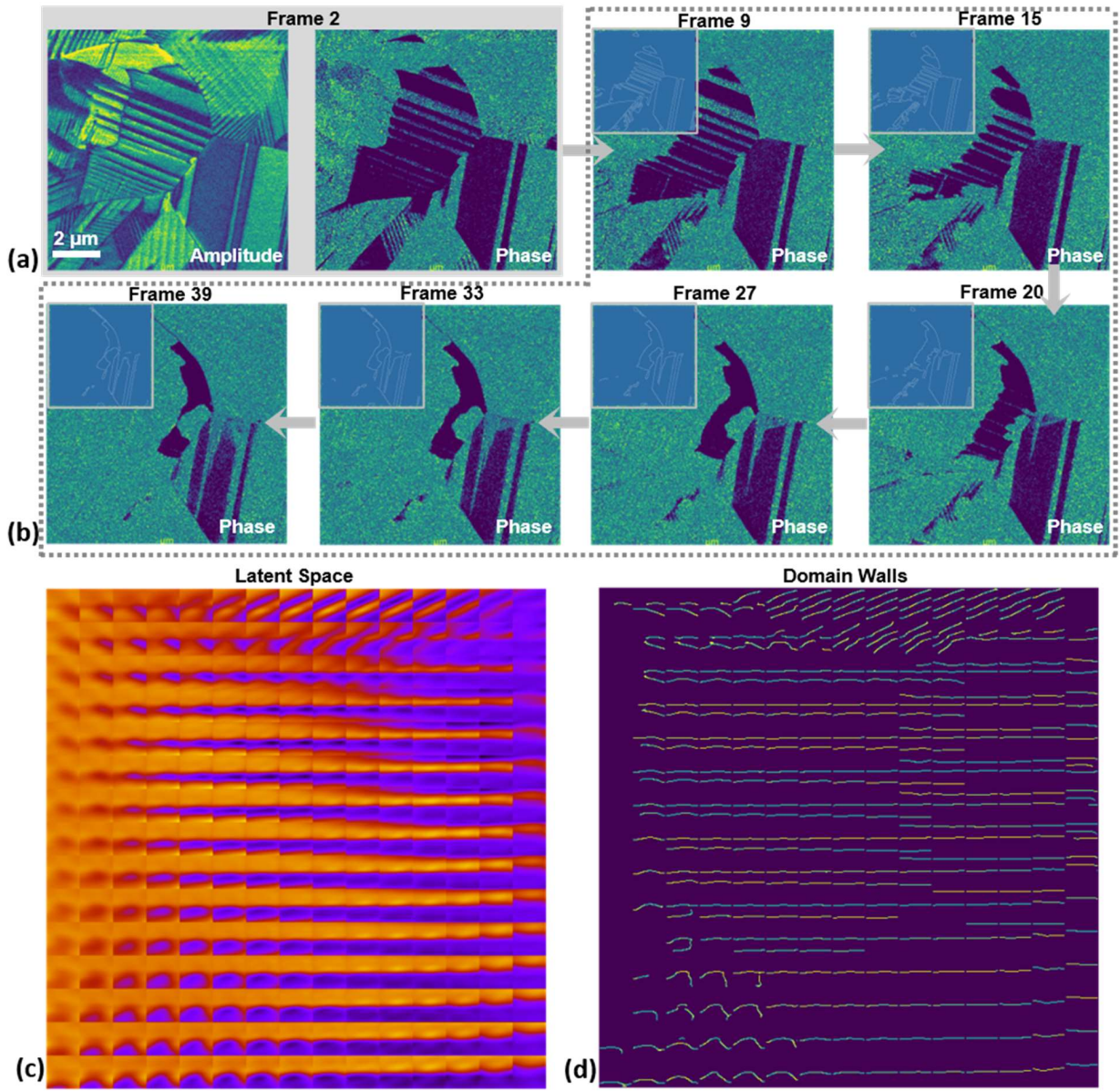

**Figure 6.** Application of skip-rVAE to piezoresponse force microscopy (PFM) data. (a) PFM

amplitude and phase image of ferroelectric domains in polycrystalline lead zirconate-titanate. (b) Time evolution of domain wall geometry during switching. Insets show corresponding domain walls extracted by Canny filter. (c) latent space representation of domain wall geometry and (d) extracted edge geometries

We further illustrate that the parsimonious discovery of physical mechanisms can be applied to other systems. Here, we use as an example the Interferometric Displacement Sensor Piezoresponse Force Microscopy (IDS-PFM)[56, 57] imaging of domain wall dynamics in lead zirconate-titanate (PZT). In IDS-PFM, the measured signal is closely related to polarization of the material and is only weakly sensitive to such factors as topography, etc.[58] that can act as a source of crosstalk,[59] or observational biases. This high veracity connection between measured signal and materials physics is common for STEM and PFM and allows direct ML analysis.

Shown in Figure 6 (a) is the typical PFM amplitude and phase images, delineating clear 180° domain wall and presence of large number of in-plane ferroelastic domains. The sequence of images in Figure 6 (b) illustrate the motion of domain wall under the action of applied bias to the probe. The motion clearly has a complex character, with the wall geometry changing due to interaction with elements of domain structure. Here, we seek to explore the associated mechanisms following the same logic as for the chemical transformations from STEM data.

To explore the dynamics of the ferroelectric domain walls, we define the descriptors as time-delayed sub-images centered at the domain wall. In other words, we identify the points at the domain wall at time $t$ and form the descriptor as a sub-image centered at this point. Note that the descriptors can be chosen at the same time, providing insight into overall domain structure, or with a time delay providing insight into associated dynamics. The corresponding latent representation is shown in Figure 6c and shows clear ordering within the latent space. Note that these latent variables encode the dynamics of the domain wall, and its evolution along the moving wall, Figure 7 (a), represents the corresponding mechanism.

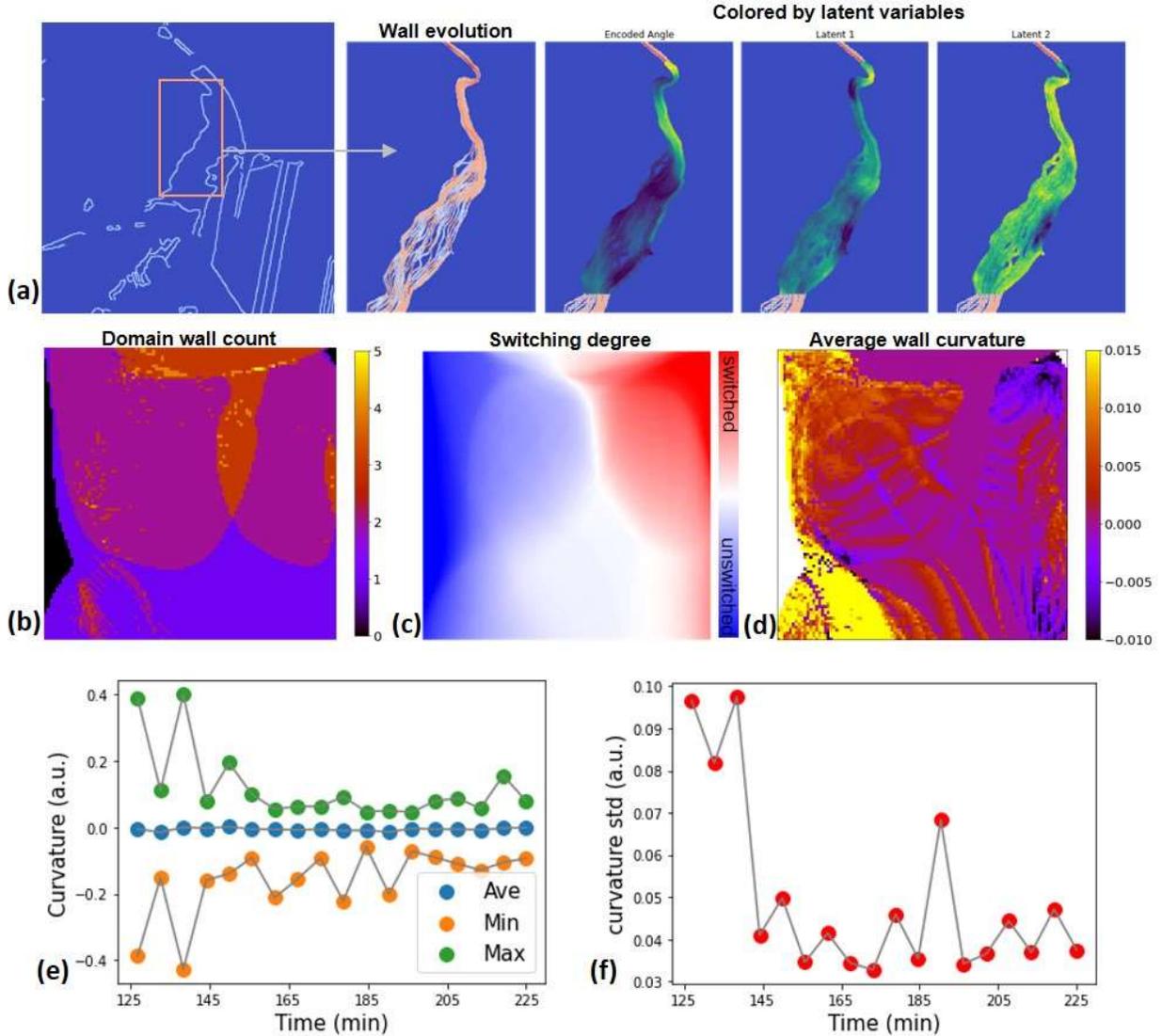

**Figure 7.** Latent space analysis of domain wall evolution. (a), Evolution of a continuous ferroelectric domain wall during switching, corresponding to frames 20-39 (the full video is available in Supplemental Material); right hand side shows the wall evolution colored by the encoded angle and latent variables. The wall images were generated by Canny filter[60] (see Methods section for details) (b) Domain switching degree, (c) domain wall count, and (d) domain wall curvature extracted from the latent space manifold. (e) Curvature evolution of the continuous domain wall shown in (a). (f) The corresponding dispersion of the wall curvature.

We further connect the mechanisms discovered via rVAE and the traditional physical descriptors. Note that analysis of lattice space illustrates that only part of it has simple structures, while another part is occupied by complex multiple wall geometries, as can be seen from Figure 6c. These regions represent evolution of needle like domains and strong wall-wall interactions. It is straightforward to introduce physics-driven parameters such as degree of switching and, for wall regions with only one or two domain walls, a wall curvature. The former descriptor reflects the local wall velocity, while the latter represents the wall bending. Accordingly, distributions of switching degree (Figure 7c) and average wall curvature (Figure 7d) indicate wall velocity and shape changes, respectively, during switching.

Note that the degree of switching shows very systematic evolution across the latent space, where zero corresponds to a static wall and positive values correspond to an advancing wall. The curvature distribution shows more complex dynamics; however, clear regions with preponderant positive and negative curvatures are visible. Note that unlike the STEM image analysis, here establishing the relationship between the latent descriptors and simple physical descriptors is possible only close to the simple wall geometries (edges), whereas latent variables can encode more complex observed mechanisms.

Finally, we can establish the evolution of latent variables during domain wall motion as shown in Figure 7a. Here, the wall encoding clearly shows complex correlated spatiotemporal structure, encoding local wall dynamics due to the presence of pinning centers. In addition, we analyzed wall curvature changes during switching (Figure 7e, f). Overall, while the average curvature is stable (Figure-7e), the difference in curvature along a continuous wall decreases during switching and thereby the curvature dispersion reduces (Figure 7f). This implies the interaction between wall shape and dynamics during ferroelectric polarization.

We have shown that unsupervised machine learning can be used to learn chemical and physical transformation pathways from observational data. In STEM, we simply assumed the existence of atoms, a discreteness of atomic class, and that there is an explicit correspondence between the observed STEM contrast and presence of atomic units. These reasonable assumptions, at the stage of the DCNN decoding of the images, enabled transitioning from the STEM data to atomic coordinates, and at the stage of latent space analysis allowed separation of physical and unphysical configurations. With only these postulates, skip-rVAE can identify the existing molecular fragments observed within the material, encode them via 2 latent variables (for convenience), and enable exploration of transformation mechanisms through tracing the evolution of atoms in the latent space of the system. Similarly, in PFM, the high-veracity relationship between the contrast and domain structure allows us to decipher the latent mechanisms of polarization dynamics.

Overall, this approach suggests that the imaging data obtained during dynamic evolution can be used to derive the chemical and physical transformation pathways involved, by providing encodings of the observed structures that act as bottom-up equivalents of structural order parameters. This in turn provides a strong stimulus towards the development of STEM, TEM, and SPM techniques, including low dose and ultrafast imaging and full information detection. Finally, we pose that a similar approach can be applicable to other atomic scale and mesoscopic imaging techniques, providing a consistent approach for the identification of order parameters and other descriptors in complex mechanisms.

Finally, we argue that this approach forms a consistent framework for application of machine learning methods in physical sciences. The necessary steps here are the clear formulation on what postulates and constraints (prior knowledge) are imposed during the feature selection and network architecture, what data is provided, and what new knowledge is revealed by the ML algorithm given the data. Here, we demonstrated the ML approach for discovery of chemical transformations from STEM observations given the existence and observability of atoms and domain wall dynamics given the existence of domain walls. Hence, we also believe that this approach will stimulate broader application of variational (i.e., Bayesian) methods in physical sciences, as well as the development of new ways to encode physical constraints in the encoder-decoder architectures, and generative physical laws and causal relationships in the latent space of VAEs.

## Acknowledgement

This effort (ML and STEM) is based upon work supported by the U.S. Department of Energy (DOE), Office of Science, Basic Energy Sciences (BES), Materials Sciences and Engineering Division (S.V.K., O.D.), by U.S. Department of Energy, Office of Science, Office of Basic Energy Sciences Data, Artificial Intelligence and Machine Learning at DOE Scientific User Facilities program under Award Number 34532 (A.G.), by the U.S. Department of Energy, Office of Science, Office of Basic Energy Sciences Energy Frontier Research Centers program under Award Number DE-SC0021118 (Y.L.), and was performed and partially supported (M.Z., B.GS.) at the Oak Ridge National Laboratory's Center for Nanophase Materials Sciences (CNMS), a U.S. Department of Energy, Office of Science User Facility operated by Oak Ridge National Laboratory. We acknowledge multiple productive interactions with Dr. Stephen Jesse.

**Methods**

*Graphene Sample Preparation*

Atmospheric pressure chemical vapor deposition (AP-CVD) was used to grow graphene on Cu foil.[61] A coating of poly(methyl methacrylate) (PMMA) was spin coated over the surface to protect the graphene and form a mechanical stabilizer during handling. Ammonium persulfate dissolved in deionized (DI) water was used to etch away the Cu foil. The remaining PMMA/graphene stack was rinsed in DI water and positioned on a TEM grid and baked on a hot plate at 150 °C for 15 min to promote adhesion between the graphene and TEM grid. After cooling, acetone was used to remove the PMMA and isopropyl alcohol was used to remove the acetone residue. The sample was dried in air and baked in an Ar-$O_2$ atmosphere (10% $O_2$) at 500 °C for 1.5 h to remove residual contamination.[62] Before examination in the STEM, the sample was baked in vacuum at 160 °C for 8 h.

*PZT sample preparation*

The measured sample was a polycrystalline lead zirconate titanate - $PbZr_{1-x}Ti_xO_3$ (PZT) with x~0.4 from Morgan Advanced Materials. The samples were cut with a diamond band saw and polished first with diamond hand polishing pads (from 50-3000 grit) and then with successively finer diamond polishing compounds, finishing with 50,000 grit lapidary paste. These materials were overstock for tube scanners in Atomic Force Microscope.

*STEM Imaging*

For imaging, a Nion UltraSTEM 200 was used, operated at 100 kV accelerating voltage with a nominal beam current of 20 pA and nominal convergence angle of 30 mrad. Images were acquired using the high angle annular dark field detector.

*PFM imaging*

The PFM imaging was performed using a commercial Cypher AFM system (Asylum Research, Santa Barbara, CA).

*STEM data analysis*

The DCNN for atomic image segmentation was based on the U-Net architecture[63] where we replaced the conventional convolutional layers in the network's bottleneck with a spatial pyramid of dilated convolutions for better results on noisy data.[64] The DCNN weights were trained using Adam optimizer[65] with cross-entropy loss function and learning rate of 0.001. In skip-rVAE, both encoder and decoder had 4 fully-connected layers with 256 neurons in each layer. The skip connections were drawn from the latent layer into every decoder layer. The latent layer had 3 neurons designated to "absorb" arbitrary rotations and *xy*-translations of images content and all the rest neurons (2 in this case) in the latent layer were used for disentangling different atomic structures. The encoder and decoder neural networks were trained jointly using the Adam optimizer with the learning rate of 0.0001 and mean-squared error loss. Both DCNN and VAE were implemented via the custom-built AtomAI package[66] utilizing PyTorch deep learning library.[67]

*PFM data analysis*

A Canny filter was applied to the PFM phase images for detecting 180° domain walls. Then, a selected number of wall pixels (80 in this analysis) was set as the center to generate sub-images with window size equal to 40 pixels. Thereby, each sub-image contains a wall located at the center. Then, the sub-images set was used for skip-rVAE-based analysis.

For the latent space analysis, the switch degree was defined as:

$$D_s = \frac{A_s - A_{us}}{A_s + A_{us}}$$

where $D_s$ is the switch degree, $A_s$ is the area of switched region, and $A_{us}$ is the area of the unswitched region. The switched area and unswitched areas are classified by the median value of latent space, the area (e.g., yellow) with values larger than the median is defined as switched area and the area (e.g., blue) is defined as unswitched area.

The wall curvature calculation was only applied to the parts containing one wall or two walls. For the parts without detectable walls, the curvature was set to NaN, which is the transparent region in the curvature map. For the parts with three or more walls, the curvature was set to zero.

For more details, see the interactive Jupyter notebooks reproducing all the data analysis available without restrictions at https://github.com/ziatdinovmax/ChemDisc